\documentclass[a4paper]{jpconf}
\usepackage{graphicx}
\newcommand{\be}{\begin{equation}}
\newcommand{\ee}{\end{equation}}

\newcommand{\bea}{\begin{eqnarray}}
\newcommand{\eea}{\end{eqnarray}}

\begin{document}
\title{Empirical Study of the GARCH model with Rational Errors}

\author{Ting Ting Chen$^{1}$ and Tetsuya Takaishi$^{2}$} 
%\author{Tetsuya Takaishi$^{2}$} 
\address{$^{1}$Faculty of Integrated Arts and Sciences, Hiroshima University, Higashi-Hiroshima 739-8521, Japan}
\address{$^{2}$Hiroshima University of Economics, Hiroshima 731-0192, JAPAN}

\ead{$^{1}$d102355@hiroshima-u.ac.jp}
\ead{$^{2}$tt-taka@hue.ac.jp}

\begin{abstract}
We use the GARCH model with a fat-tailed error distribution described by 
a rational function and apply it for the stock price data on the Tokyo Stock Exchange. 
To determine the model parameters we perform the Bayesian inference to the model.
The Bayesian inference is implemented by the Metropolis-Hastings algorithm
with an adaptive multi-dimensional Student's t-proposal density.
In order to compare the model with the GARCH model with the standard normal errors 
we calculate information criterions: AIC and DIC, 
and find that both criterions favor the GARCH model with a rational error distribution.
We also calculate the accuracy of the volatility by using the realized volatility
and find that a good accuracy is obtained for  the GARCH model with a rational error distribution. 
Thus we conclude that the GARCH model  with a rational error distribution
is superior to the GARCH model with the normal errors and
it can be used as an alternative GARCH model to those with other fat-tailed distributions.
\end{abstract}

\section{Introduction}
In finance volatility plays a central role for risk management such as 
derivative price estimation and portfolio allocation for which it is important
to measure a reliable volatility from the data observed in the financial markets.
Since volatility is not a direct observable in the financial markets
we have to rely on  a certain estimation technique.
Usually parametric volatility models designed to
capture asset return and volatility properties are often used
in empirical finance. 
The most popular and successful model is the GARCH model\cite{GARCH}
which is a generalized version of the ARCH model\cite{ARCH}.

In the GARCH model asset returns $r_t$ at time $t$ are expressed as  
$r_t=\sigma_t \epsilon_t$ where $\sigma_t^2$ is the time-changing volatility which is given by 
a function of past returns and  past volatilities. 
In the original GARCH model the standard normal distribution, i.e. iid $N(0,1)$ 
was used for $\epsilon_t$ errors. 
It is known that the GARCH model well captures
relevant properties of asset returns
such as the fat-tailed behavior of the return distribution and 
the volatility clustering that are classified as the stylized facts\cite{Cont}.
On the other hand in empirical studies
it is often observed that the GARCH model does not sufficiently account for 
the leptokurtosis of the return distributions. 
To circumvent this 
it is advocated to apply a different distribution having a fatter tail than that of 
the normal distribution for the $\epsilon_t$ error.
Several distributional forms having a fatter tail than the normal distribution
such as Student's t-distribution\cite{GARCH2} 
and the generalized error distribution (GED)\cite{GED} are  
applied for the $\epsilon_t$ error term.
By using the Student's t-distributions  or GED for $\epsilon_t$ errors usually one gets 
a better goodness-of-fit to the financial return data.
However the Student's t-distributions  and GED  are not necessarily the optimal solution for  the $\epsilon_t$ error
term of the GARCH model
and  one could also choose other fat-tailed distributions. 

In this study we apply  Pad\'e approximants described 
by a rational function for the $\epsilon_t$ error term.
The Pad\'e approximants are flexible to approximate a function in a certain domain. 
In finance Pad\'e approximants are used to describe
the interest rate return distributions\cite{Nuyts,Nuyts2},
where the parameters of rational functions are obtained by fitting to the interest rate return distributions. 
Here we apply a rational function for the  $\epsilon_t$ error of the GARCH model.
In Ref.\cite{GARCHRE}
the GARCH model with rational errors
was investigated by using  USD/JPN exchange rate returns and
the goodness-of-fit by Akaike information criterion (AIC)\cite{AIC} 
and deviance information criterion (DIC)\cite{DIC} 
showed that the GARCH model with rational errors is superior to the GARCH mode with normal errors. 

We further investigate the effectiveness of  the GARCH model with rational errors 
by using stock return data on the Tokyo Stock Exchange.
In this study in order to clarify the model-effectiveness, 
in addition to AIC and  DIC, we 
utilize realized volatility which is a model-free estimate of 
the integrated volatility. 
Using realized volatility as a proxy of the true volatility 
we calculate the accuracy of the volatility by a loss function for both models
and 
using the loss function we compare which model is more effective.

\section{GARCH model with normal error distribution} 
Bollerslev introduced the GARCH(p,q) model\cite{GARCH} which is 
a generalized version of the ARCH model\cite{ARCH}.
The GARCH(p,q) model is expressed as
\be
y_t=\sigma_t \epsilon_t ,
\ee
and 
\be
\sigma_t^2  = \omega + \sum_{i=1}^{q}\alpha_i y_{t-i}^2
+ \sum_{i=1}^{p}\beta_i \sigma_{t-i}^2,
\ee
where $\alpha_i$, $\beta_i$ and $\omega$ are parameters of the GARCH model.
These parameters are determined so that the model matches the return data.
Since the volatility $\sigma_t^2$ should be positive
the GARCH parameters are restricted to $\omega>0$, $\alpha_i>0$ and $\beta_i>0$ to ensure a positive volatility.
%and the stationary condition $\sum_{i=1}^{q}\alpha_i + \sum_{i=1}^{p}\beta_i <1$ is also required.
$\epsilon_t$ is an independent normal error following $N(0,1)$ 
and the return time series is given by $y_t$.
In this study 
we focus on the GARCH(1,1) model, i.e. $p=1$ and $q=1$, 
where the volatility process is given by
\be
\sigma_t^2  = \omega + \alpha y_{t-1}^2 + \beta \sigma_{t-1}^2,
\ee
and hereafter the GARCH model simply denotes the GARCH(1,1) model. 
Moreover for the GARCH model with normal errors 
we call it  the GARCH-N model.

\section{GARCH model with rational error distribution}
In general a rational function of the Pad\'e approximants is expressed by
two polynomial functions $T_M(x)$ and $B_N(x)$ as
\be
P_{M,N}(x)=\frac{T_M(x)}{B_N(x)},
\ee
where $M$ and $N$ stand for the degrees of the polynomial  $T_M(x)$ and $B_N(x)$ 
respectively.
In order to consider $P_{M,N}(x)$ as a probability distribution 
we have to impose  conditions that it must be positive and normalized to 1.
Furthermore similar to the normal distribution we assume that  $P_{M,N}(x)$  
takes a maximum value at the origin and it is symmetric to the $x=0$ axis, i.e. $P_{M,N}(x)=P_{M,N}(-x)$.
In Ref.\cite{Nuyts} possible normalizable distributions with finite variances are derived to approximate the interest rate distributions. 
The simplest normalized probability distribution with tunable parameters $a_1$ and $a_2$
is given by
\be
P_{0,4}(x)=\frac{a_1}{\pi(1+(a_1^2+2a_2)x^2+a_2^2x^4)}.
\ee
The variance of this probability distribution is calculated to be $-1/a_2$.
Since usually the variance of the GARCH error distribution is set to 1
we also set the variance of $P_{0,4}(x)$ to 1, i.e. $a_2=-1$.
Finally we obtain our rational error distribution for the GARCH model  as
\be
P(x)=\frac{a}{\pi(1+(a^2-2)x^2+x^4)}.
\label{eq:R}
\ee
When we use the rational error distribution of (\ref{eq:R}) for $\epsilon_t$ of the GARCH model
we call it the GARCH model with rational errors (GARCH-RE model)\cite{GARCHRE}.

%\section{Bayesian estimation of the GARCH parameters}

\section{Realized volatility}
Recent availability of high frequency financial data 
enables us to calculate the realized volatility 
constructed as a sum of squared intraday returns\cite{RV1,RVGaussian,RV2,RV3}, see also e.g.\cite{RVRev}.
Let us assume that the logarithmic price process $\ln p(s)$ follows
a continuous time stochastic diffusion,
\be
d\ln p(s) =\tilde{\sigma}(s)dW(s),
\label{eq:SD}
\ee
where $W(s)$ stands for a standard Brownian motion  and $\tilde{\sigma}(s)$ is a spot volatility at time $s$.
Under this assumption the integrated volatility defined by
\be
\sigma_h^2(t) =\int_{t}^{t+h}\tilde{\sigma}(s)^2ds,
\label{eq:int}
\ee
where $h$ stands for the interval to be integrated.
In empirical finance "daily volatility" is of primary importance and 
for the daily volatility  $h$ takes one day.
Since $\tilde{\sigma}(s)$ is latent and not observed in the financial markets,
(\ref{eq:int}) can not be evaluated analytically.

Let us define a sampling period $\Delta$ by $\Delta=h/n$, i.e.
we sample $n$ returns in the time interval of $h$.
Then the $i-th$ intraday return on the day $t$ with $\Delta$ sampling period is given by a log-price difference as
\be
r_{t+i\Delta}=\ln P_{t+i\Delta}-\ln P_{t+(i-1)\Delta},
\ee
where $P_t$ is an asset price at time $t$.
Using these intraday returns
the realized volatility $RV_t$ on the day $t$ is given by 
a sum of squared intraday returns as
\be
RV_t =\sum_{i=1}^{n} r_{t+i\Delta}^2.
\label{eq:RV}
\ee

Under ideal circumstance 
$RV_t$ is proved to converge to the integrated volatility of (\ref{eq:int}) in the limit of $n \rightarrow \infty$.
However in the real financial markets 
there exist several types of bias such as microstructure noise\cite{Campbell},
and thus in the presence of the bias 
the convergence of $RV_t$ to the integrated volatility is not guaranteed.
Let us assume that the log-price observed in financial markets 
is contaminated with independent noise\cite{Zhou}, i.e.
\be
\ln P^{*}_t =  \ln P_t +\xi_t,
\ee
where $\ln P_t^{*}$ is the observed log-price in the markets which consists of the true log-price $\ln P_t$ and
noise $\xi_t$ with mean 0 and variance $\rho^2$.
Under this assumption the observed
return $r^{*}_t$ is given by
\be
r^{*}_t=r_t +\eta_t,
\ee
where $\eta_t=\xi_{t}-\xi_{t -\Delta}$.
Thus $RV^{*}_t$ actually observed from the market data is obtained
as a sum of the squared returns $r^{*}_t$,
\bea
RV_t^{*}& = &\sum_{i=1}^{n} (r^{*}_{t+i\Delta})^2,  \\
   & =& RV_t + 2\sum_{i=1}^n r_{t+i\Delta}\eta_{t+i\Delta} + \sum_{i=1}^n \eta_{t+i\Delta}^2.
\label{eq:rvnoise}
\eea
With these independent noises  the bias appears as $\sum_{i=1}^n \eta_{t+i\Delta}^2$
which corresponds to $\sim 2n\rho^2$.
Thus due to the bias the $RV^{*}_t$ diverges as $n \rightarrow \infty$.

Practically in order to avoid the distortion from microstructure noise 
one needs to choose a good sampling period  which
reduces the microstructure noise bias  and at the same time to maintain the accuracy of the realized volatility.
The optimal sampling period is suggested to be around 5min\cite{Bandi}.
One could also use kernel-based estimations which are designed to reduce the microstructure noise\cite{Zhou,HL1,BN1}.

Another type of bias is due to "non-trading hours".
Since stock markets are not open 24 hours
the high-frequency data are only available for a part of 24 hours.
At the Tokyo stock exchange market domestic stocks are traded in
the two trading sessions: (a) morning trading session 9:00-11:00.
(b) afternoon trading session 12:30-15:00.
The daily realized volatility calculated without including intraday returns during
the non-trading periods
can be underestimated.
When we consider volatility only in each trading session\cite{TakaishiRV,TakaishiRV2}
this bias problem does not arise.
Otherwise we need to deal with this bias appropriately. 

Hansen and Lunde\cite{Hansen} advocated an idea to circumvent the problem by
introducing an adjustment factor which
modifies the realized volatility so that the average of the realized volatility matches
the variance of the daily returns.
Let $(R_1,...,R_N)$ be $N$ daily returns constructed by 
close-close daily log-price difference.
The adjustment factor $c$ ( HL adjustment factor)  is given by
\be
c=\frac{\sum_{t=1}^{N}(R_t-\bar{R})^2}{\sum_{t=1}^{N}RV_t},
\ee
where $\bar{R}$ denotes the average of $R_t$.
Then using this factor the daily realized volatility is modified to $cRV_t$.
Although originally the HL adjustment factor is introduced to correct the bias of the non-trading hours
it can also correct the microstructure noise bias effects to some extent.

\section{Empirical results}

\begin{figure}[ht]
\vspace{1cm}
%\begin{miniprre}{14pc}
\begin{center}
\includegraphics[width=23pc]{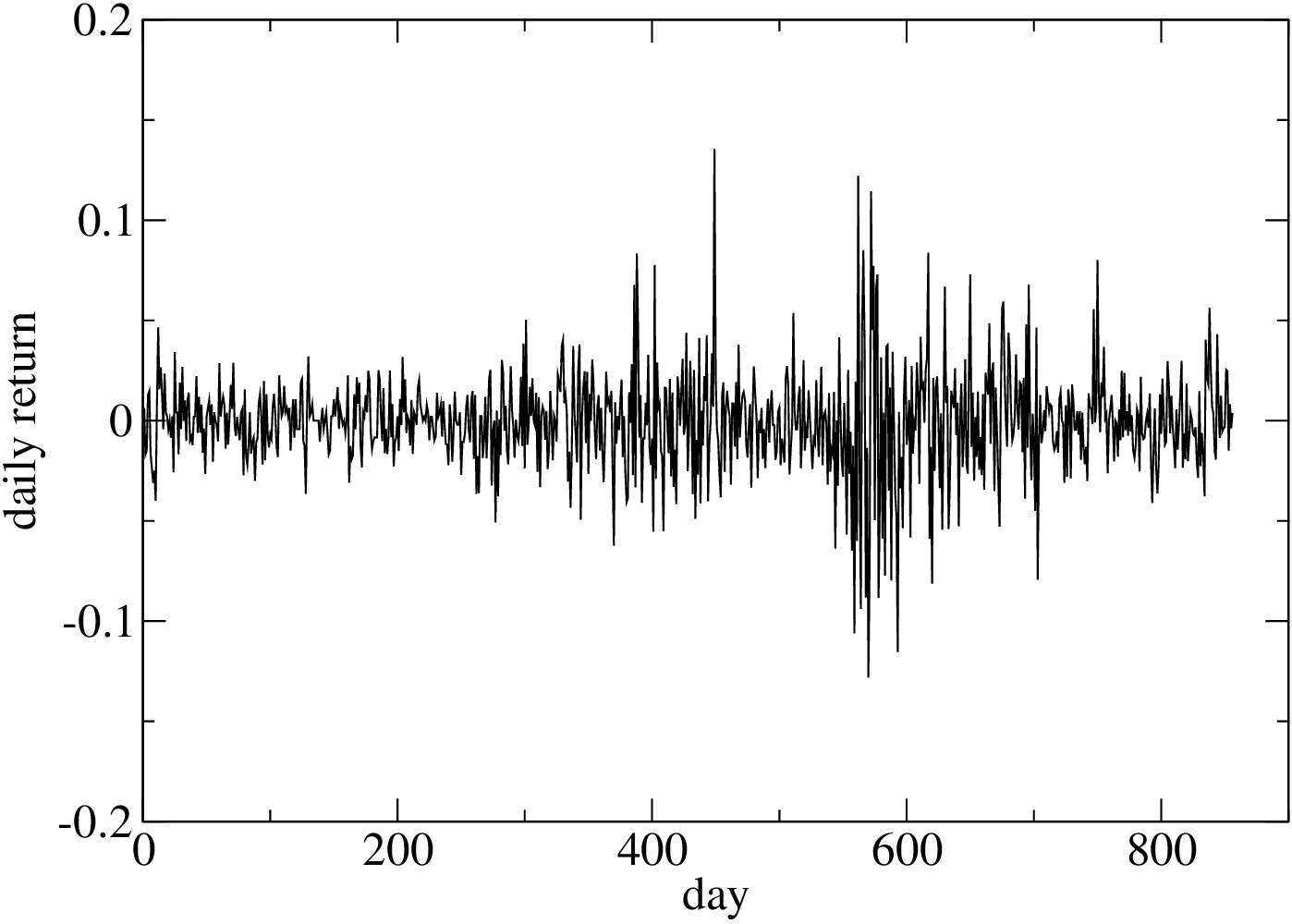}
\end{center}
\caption{
Daily return time series of Panasonic Co.}
%\end{minipage}
%\hspace{5pc}%
\end{figure}

In this study we analyze the stock price data of Panasonic Co. traded on the Tokyo Stock Exchange.
This stock is listed in the Topix core 30 index which includes the 30 most liquid and highly market capitalized stocks. 
Our data set begins June 3, 2006 and ends December 30, 2009.
Figure 1 shows the daily return time series of Panasonic Co.
We apply the GARCH-RE and GARCH-N models for the daily returns shown in Figure 1 
and estimate the daily volatilities corresponding to those daily returns. 
The parameter estimation of the GARCH-RE and GARCH-N models is conducted by the Bayesian inference.
A popular approach to perform the the Bayesian inference is the Markov Chain Monte Carlo (MCMC) methods.
Since there exist a variety of MCMC methods we need to choose an adequate method for the Bayesian inference of the GARCH model.
We perform the Bayesian inference by the Metropolis-Hastings algorithm\cite{Metropolis,Hastings} 
with an adaptive multi-dimensional Student's t-proposal density (MHAS algorithm)\cite{Takaishi1,Takaishi2,Takaishi3,Takaishi4}.
In the MH algorithm we need to specify the proposal density.
In Ref.\cite{Nakatsuma} the proposal densities constructed from an auxiliary process are used for the MH algorithm.
Refs.\cite{Watanabe,Asai} use a multi-dimensional Student's t-proposal density
for which density parameters are determined by the maximum likelihood method. 
Here we use MHAS algorithm where density parameters of a multi-dimensional Student's t-proposal density
are determined adaptively during the Monte Carlo simulations so that 
the multi-dimensional Student's t-proposal density matches the posterior distributions of the model.
The MHAS algorithm has been shown to be very efficient 
for the Bayesian inference of the GARCH models\cite{Takaishi1,Takaishi2,Takaishi3,Takaishi4}. 
The implementation of the MHAS algorithm was done as follows.
We discarded the first 6000 Monte Carlo updates by the MHAS algorithm.
Then we accumulated  50000 Monte Carlo samples for analysis.
Table 1 shows the values of the parameters averaged over the Monte Carlo samples.
The values marked by $*$ show the autocorrelation time of the Monte Carlo data generated 
by the MHAS algorithm.
We find that the values of the autocorrelation time are small which indicates that
the MHAS algorithm generates effectively un-correlated Monte Carlo samples.

In order to compare the goodness-of-fit of the models
we utilize  two information criterions: AIC\cite{AIC} and
DIC\cite{DIC}.
The AIC is defined by
$AIC=-\ln L(\bar{\theta})-2k$
where $k$ is the number of the parameters of the model and
$L(\bar{\theta})$ is the likelihood function of the model at $\bar{\theta}$.
$\theta$ stands for $\theta=(\alpha,\beta,\omega,a)$ for the GARCH-RE model and  $\theta=(\alpha,\beta,\omega)$ for the GARCH-N model.
$\bar{\theta}$ stands for the parameters averaged over the Monte Carlo samples.
The DIC is defined by $2[\ln L(\bar{\theta})-2E(\ln L(\theta))]$
where $E(\ln L(\theta))$ is the Monte Carlo average of $\ln L(\theta)$.  
For both AIC and DIC  the model with the smallest value is chosen as the 
one which would best predict the time series. 
As seen in Table 1 both of AIC and  DIC give smaller values for the GARCH-RE model. 
Thus  we find that the GARCH-RE model is superior to the GARCH-N model.

\begin{table}
\caption{Results of the Bayesian inference for GARCH-RE and GARCH-N models.
The values marked by $*$ show the autocorrelation time $\tau_{int}$ defined by
$\tau_{int}=1 + 2\sum_{t=1}^{\infty} ACF(t)$, where $ACF(t)$ stands for the autocorrelation function.}
\center{
\begin{tabular}{c|cc|cc} \hline
            &    \multicolumn{2}{c}{GARCH-RE}           &  \multicolumn{2}{|c}{GARCH-N}          \\ \hline
$\alpha$    &   0.132(38)     &    $ 5.4(10)^*$         &  0.148(31)  & $2.7(3)^*$              \\  
$\beta$     &   0.858(41)     &   $5.8(12)^*$           &  0.836(33)  & $2.8(2)^*$              \\
$\omega$    &   $2.8(1.2)\times 10^{-5}$ & $6.5(16)^*$ & $1.3(5)\times 10^{-5}$ & $3.3(4)^*$   \\
$a$         &  1.57(9)        & $4.2(6)^*$              &    --       & --                       \\   \hline
AIC         &    \multicolumn{2}{c}{-4151.29}           &  \multicolumn{2}{|c}{-4148.35}                \\
DIC         &    \multicolumn{2}{c}{-4156.30}           &  \multicolumn{2}{|c}{-4151.98}                \\ \hline
\end{tabular}
}
\end{table}

\begin{figure}[]
\vspace{1cm}
%\begin{miniprre}{14pc}
\begin{center}
\includegraphics[width=23pc]{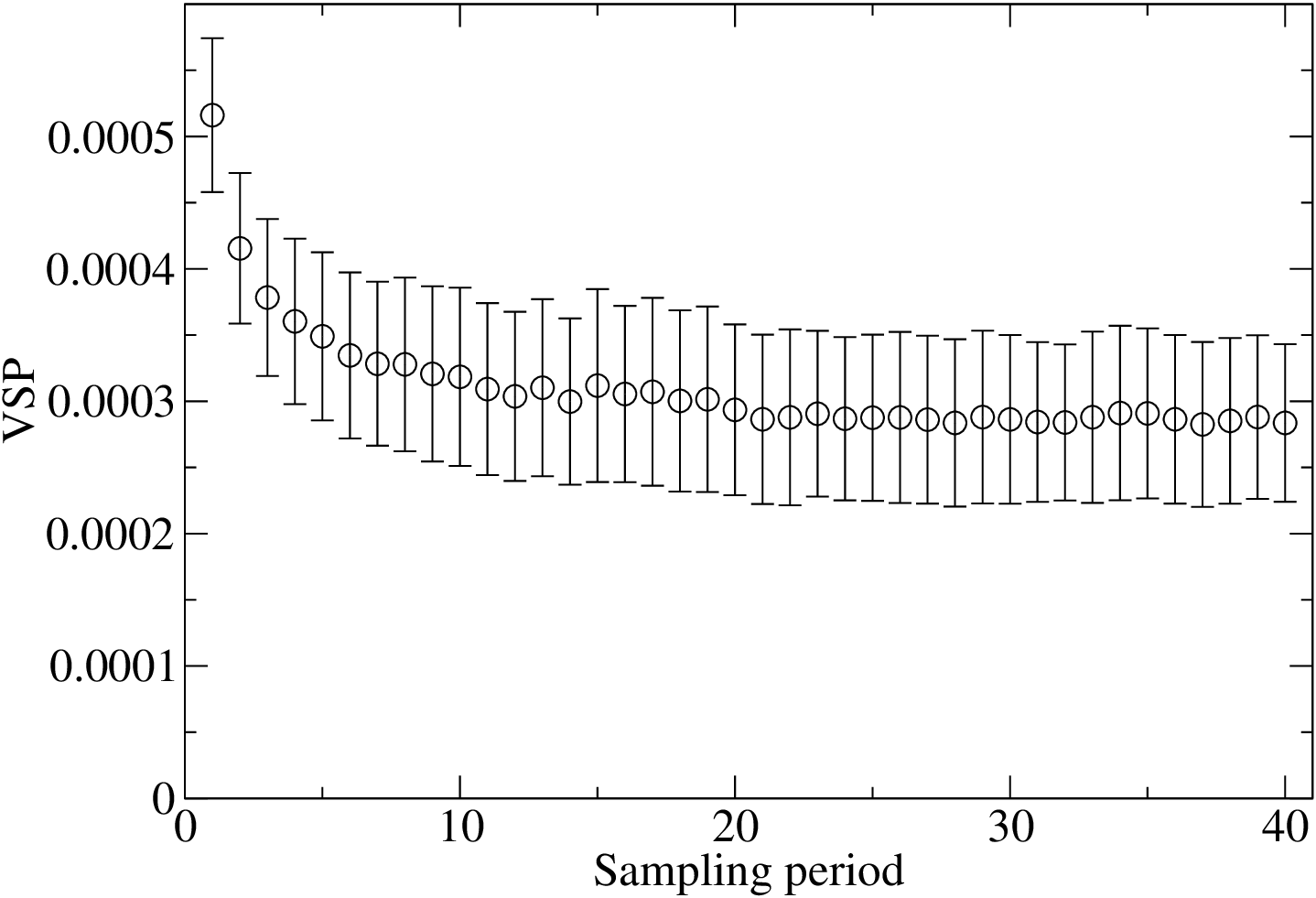}
\end{center}
\caption{
Volatility signature plot: Average realized volatility at each sampling period.}
%\end{minipage}
%\hspace{5pc}%
\end{figure}

\begin{figure}[]
\vspace{1cm}
%\begin{miniprre}{14pc}
\begin{center}
\includegraphics[width=23pc]{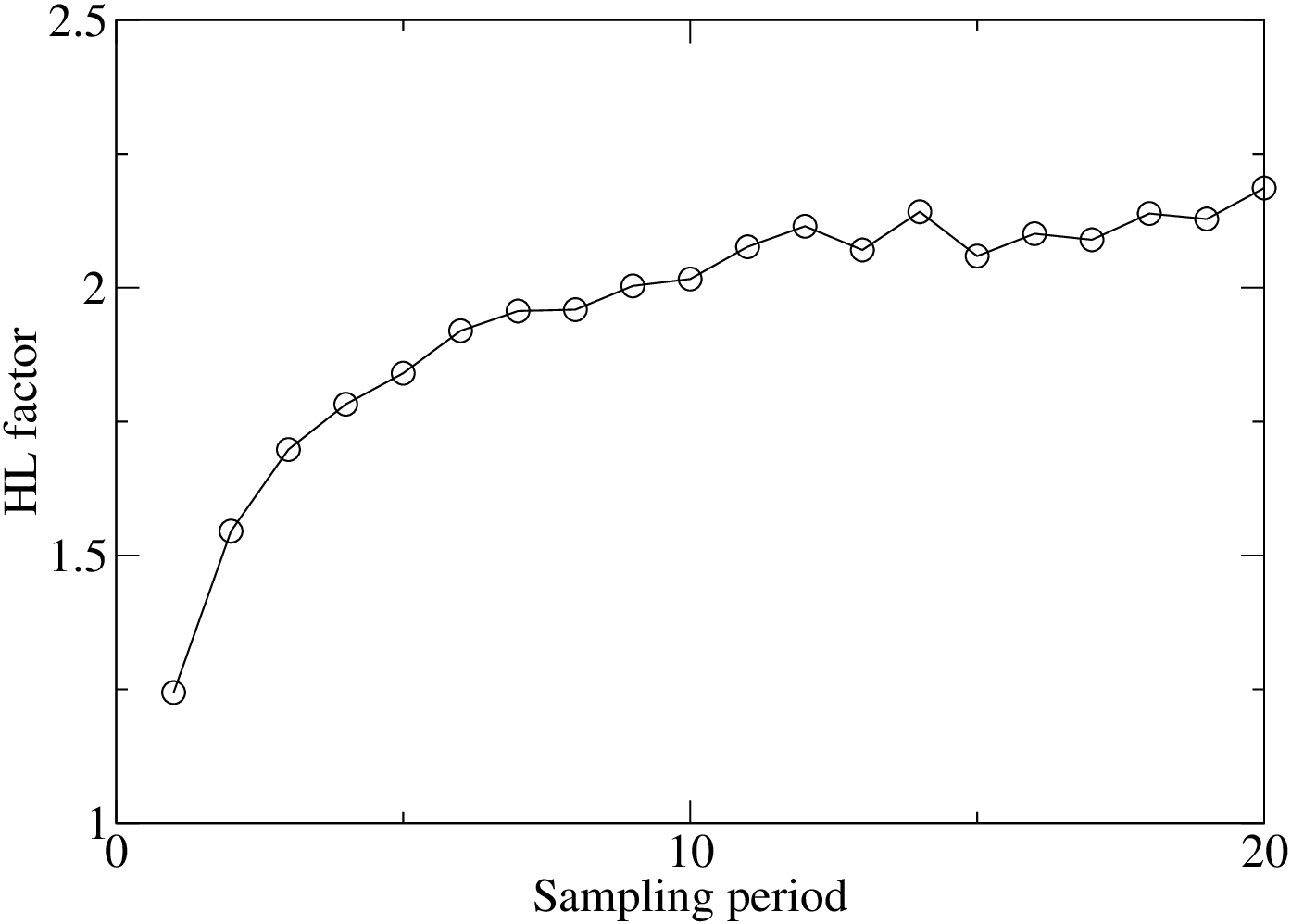}
\end{center}
\caption{
HL adjustment factor as a function of sampling period.}
%\end{minipage}
%\hspace{5pc}%
\end{figure}

\begin{figure}[]
\vspace{0.5cm}
%\begin{miniprre}{14pc}
\begin{center}
\includegraphics[width=23pc]{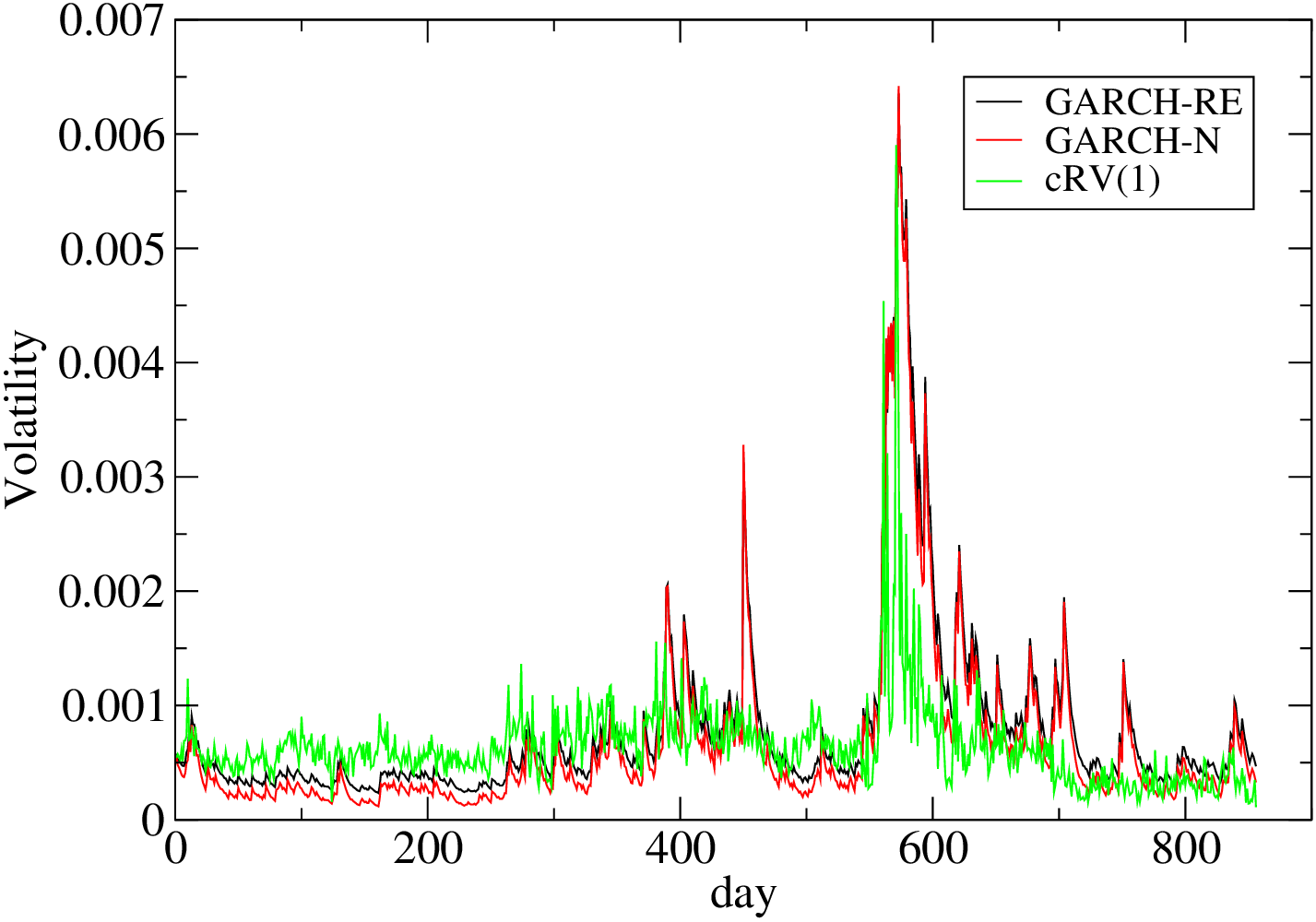}
\end{center}
\caption{
Volatility from GARCH-RE, GARCH-N models and the realized volatility at 1-min sampling period.
The realized volatility is adjusted by the HL adjustment factor.}
%\end{minipage}
%\hspace{5pc}%
\end{figure}

Next we compare the GARCH-RE and GARCH-N models with the accuracy of the volatility. 
To do that we measure the difference between the volatility from the models
and the true volatility.
Since we do not know the value of the true volatility 
we use the realized volatility as a proxy of the true volatility. 
The realized volatility is constructed by a sum of squared intraday returns as (\ref{eq:RV}). 
Figure 2 shows the average realized volatility as a function of sampling period, i.e.
the realized volatility is averaged at each sampling period.
Such plot is call "volatility signature plot" advocated in Ref.\cite{VSP}
to visualize the microstructure noise bias on the realized volatility.
As expected in  (\ref{eq:rvnoise}) we find  that 
the realized volatility diverges at small sampling period ( or at high sampling frequency ). 
In this study we correct this bias with the HL adjustment factor which adjusts the average of the realized volatility to
the variance of the daily return.

Figure 3 shows the HL adjustment factor as a function of sampling period.
The HL adjustment factor decreases as the sampling period decreases.
This decrease is explained by the microstructure noise bias which inflates the realized volatility at small sampling periods.
As the sampling period increases the HL adjustment factor reaches a plateau around 2 
where the microstructure noise bias effects are expected to be small.
This factor of 2 means that the original realized volatility is underestimated due to non-trading hours and 
the size of the volatility during no-trading hours is about the same size as that during the trading hours.
On the Tokyo Stock Exchange there are two non-trading periods: lunch break and night break.
Since during the lunch break the size of the volatility is observed to be small\cite{TakaishiRV} 
the dominant contribution to the factor of 2 comes from the night break.

\begin{figure}[]
\vspace{1cm}
%\begin{miniprre}{14pc}
\begin{center}
\includegraphics[width=23pc]{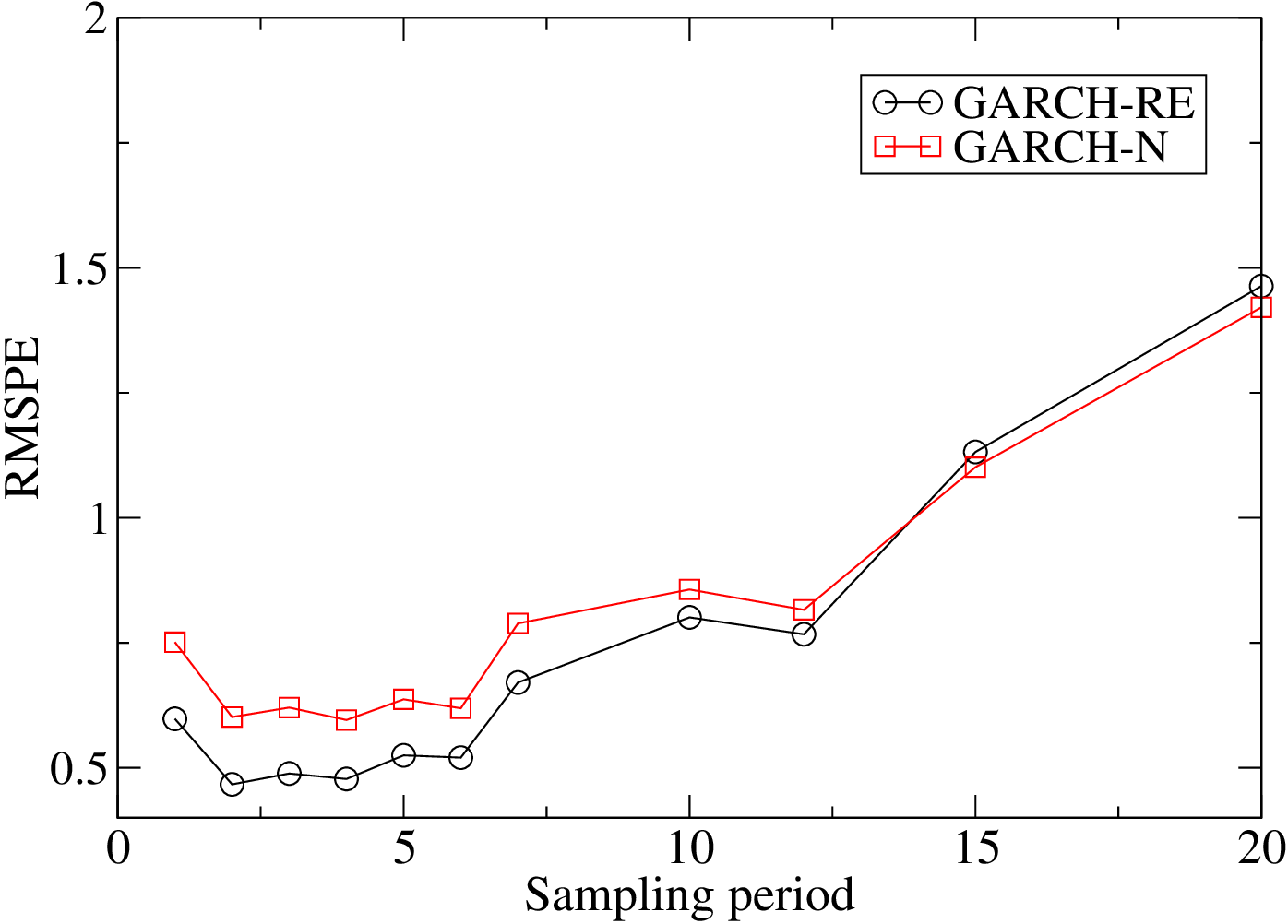}
\end{center}
\caption{
RMSPE of GARCH-RE and GARCH-N models as a function of sampling period.}
%\end{minipage}
%\hspace{5pc}%
\end{figure}

Figure 4 compares volatilities from the GARCH-RE and GARCH-N models, and the realized volatility at 1-min sampling period.
To quantify the accuracy of the volatility 
we measure a loss function of the root mean square percentage error ( RMSPE ) 
defined by 
\be
RMSPE= \left (\sum_{t=1}^N (\frac{\bar{\sigma_t^2}-cRV_t}{cRV_t})^2 \right)^{1/2},
\ee
where 
$\bar{\sigma_t^2}$ is the volatility estimated from the Bayesian inference of the GARCH-RE or GARCH-N models.
$\bar{\sigma_t^2}$ is also adjusted so that the average of $\bar{\sigma_t^2}$, i.e. $\sum_{t=1}^N \bar{\sigma_t^2}/N $ matches 
the variance of the daily returns.
Figure 5 shows RMSPE of the GARCH-RE and GARCH-N models.
We find that RMSPE takes a minimum around 1 to 6-min sampling periods where 
the GARCH-RE model gives smaller values. 
It is also noted that the sampling frequencies which take 
the minimum of the RMSPE are very similar to the optimum sampling frequencies 
obtained from the mean squared error of the realized volatility\cite{Bandi}. 
Our result of the RMSPE also indicates that the GARCH-RE  model is more effective than the GARCH-N model.

\section{Conclusions}
We performed the Bayesian inference of the GARCH-RE model and the GARCH-N model for the stock price data of Panasonic Co.  
on the Tokyo Stock Exchange. 
The Bayesian inference is implemented by the MHAS algorithm. 
In order to compare models 
we calculate information criterions: AIC and DIC, and find that both criterions favor the GARCH-RE model. 
We also calculate the accuracy of the volatility by the RMSPE
and find that the smaller RMSPE is obtained for the GARCH-RE model.
Thus we conclude that the GARCH-RE model is superior to the GARCH-N model and 
it can be used as an alternative GARCH model to those with other fat-tailed distributions. 

\section*{Acknowledgments}
Numerical calculations in this work were carried out at the
Yukawa Institute Computer Facility
and at the facilities of the Institute of Statistical Mathematics.
This work was supported by Grant-in-Aid for Scientific Research (C) (No.22500267).

\end{document}